\documentclass{aa}  

\usepackage{graphicx}

\usepackage{multirow}

\usepackage{fixltx2e}
 
\usepackage[varg]{txfonts}

\usepackage{array}

\usepackage[retainorgcmds]{IEEEtrantools}
\IEEEeqnarraydefcolsep{0}{\leftmargini}

\newcolumntype{x}[1]{>{\raggedleft\hspace{0pt}}p{#1}}
 
\usepackage{natbib}
\bibpunct{(}{)}{;}{a}{}{,} 
 
\begin{document}

\title{Understanding angular momentum transport in red giants: \\
  the case of KIC~7341231}

\author{
  T. Ceillier \inst{1}
  \and
  P. Eggenberger \inst{2}
  \and
  R. A. Garc\'ia \inst{1}
  \and
  S. Mathis \inst{1}
}

\institute{
  Laboratoire AIM Paris-Saclay, CEA/DSM/IRFU/SAp - CNRS - Universit\'e
  Paris Diderot, Centre de Saclay, F-91191 Gif-sur-Yvette Cedex, France
  \and
  Observatoire de Gen\`eve, Universit\'e de Gen\`eve,
  51 chemin des Maillettes, CH-1290, Sauverny, Suisse
}

\date{Submitted 14 March 2013, Accepted 22 May 2013}


\abstract
{Thanks to recent asteroseismic observations, it has been possible to
  infer the radial differential rotation profile of subgiants and red giants.}
{We want to reproduce through modeling the observed
  rotation profile of the early red giant KIC~7341231 and constrain
  the physical mechanisms responsible for angular momentum transport
  in stellar interiors.}
{We compute models of KIC~7341231 including a treatment of shellular
  rotation and we compare the rotation profiles obtained with the one
  derived by \citet{Deheuvels2012}. We then modify some modeling
  parameters in order to quantify their effect on the obtained
  rotation profile. Moreover, we mimic a powerful angular momentum
  transport during the Main Sequence and study its effect on the
  evolution of the rotation profile during the subgiant and red giant
  phases.}
{We show that meridional circulation and shear mixing alone produce a
  rotation profile for KIC~7341231 too steep compared to the observed
  one. An additional mechanism is then needed to increase the internal
  transport of angular momentum. We find that this undetermined
  mechanism has to be efficient not only during the Main Sequence but
  also during the much quicker subgiant phase. Moreover, we point out
  the importance of studying the whole rotational history of a star in
  order to explain its rotation profile during the red giant
  evolution.}
{}


\maketitle

%

\section{Introduction}

Stellar evolution is a very complex process which is not yet fully
understood. It is driven by many different physical mechanisms among
which rotation is known to be playing a very important part
\citep[e.g.][]{Pinsonneault1990, Eggenberger2010b, Maeder2009}. The
transport of chemical elements and angular momentum by the large-scale
meridional circulation and shear mixing, both induced by rotation, can
strongly modify the global evolution of a star
\citep{1992A&A...265..115Z, 1998A&A...334.1000M, Mathis2004}. That is
why this transport has been included in stellar evolution codes. This
implementation follows the shellular rotation hypothesis, which
assumes that a strong horizontal turbulent transport in stably
stratified stellar radiation zones leads to an angular velocity
constant on the isobars and thus depending almost only on the radius
\citep{2000A&A...361..101M, Decressin2009}. In order to validate these
models, it is now necessary to compare their results to
observations.

The first constraints that can be used to evaluate the efficiency of
rotational mixing in stars are the surface abundances. The influence
of rotation has indeed been studied on light elements abundances 
\citep[e.g.][]{Pinsonneault2010} and on the evolution of the lithium depletion 
\citep[e.g.][]{Eggenberger2012a}. The fact that the models including shellular
rotation are unable to reproduce lithium abundance in certain cases
\citep{Talon2003} is a first clue that an additional physical
process is at work in the internal layers of low-mass stars
\citep{Charbonnel2005}.

Another way to compare the results of modeling on angular momentum
transport with reality is to look directly at observations of rotation
in stars. The more straightforward way of doing so is to study
measurements of surface rotation. Such studies tend to show that young
solar-type stars in rapid rotation would demonstrate a solid-body
rotation \citep[see for instance][]{Denissenkov2010}, sign of a
strong coupling between the core and the envelope which is not
accounted for by meridional circulation and shear mixing alone
\citep{Eggenberger2010}. This is yet another indication that models
are missing another transport and mixing process.

One can then go deeper into stars by means of helio- and
asteroseimology. The study of stellar oscillations is indeed used to
probe their internal layers. As for the Sun, the vast amount of
seismological observations have allowed to derive its rotation profile
from the surface to the most inner parts \citep[e.g.][]{Mathur2008,
  2007Sci...316.1591G, Thompson1996, Elsworth1995}.  The resulting
profile is almost flat down to $0.2\ \text{R}_\odot$, while solar
models tend to produce a steeper profile
\citep[e.g.][]{2010ApJ...715.1539T, 1989ApJ...338..424P}, showing once
again the need for an additional coupling process.

To this day, two possible mechanisms have been investigated. The first
one implies fossil magnetic fields trapped during the early phases of
the star evolution \citep{Braithwaite2004, Duez2010}. These magnetic
fields are able to transport angular momentum thanks to Maxwell
stresses, large-scale torques and magnetic instabilities
\citep[e.g.][]{Dough1998, 2005A&A...440..653M, Garaud2008,
  2011A&A...532A..34S, Eggenberger2005}. The second one consists in
internal gravity waves, excited at the limit between the convective
and radiative zones. Such waves can indeed transport angular momentum
through radiative zones \citep[e.g.][]{Press1981, Goldreich1989,
  1993A&A...279..431S, Zahn1997, Talon2005, Mathis2012}.

The launch of space missions such as CoRoT \citep{2006ESASP1306...33B}
and \emph{Kepler} \citep{Borucki19022010} have given access to long
high-precision light curves for a huge number of stars. Many of these
stars exhibit solar-like oscillations at different evolutionary
stages, from the main sequence \citep[see for
instance][]{2008A&A...488..705A, Garcia2009b, Chaplin2011, Ballot2011,
  Metcalfe2012, Mathur2013} to more advanced stages such as subgiants
\citep[among others]{Mathur2011, Campante2011} and red giants
\citep[e.g.][]{Huber2011, Bedding2010, Mosser2010a,
  Mosser2012,DiMauro2011} which allowed to probe stellar
interiors. Moreover, the lenght of the time series available is key to
study the dynamics of stars. The existence of surface magnetic
activity, for instance, has made possible to constrain the surface
rotation \citep{Mathur2010, Frohlich2012a} as well as to measure
activity cycles \citep{Garcia2010b}.

Furthermore, the precise analysis of the rotational splittings of the
oscillations modes can strongly constrain the internal rotation
profile of main sequence stars \citep{Ballot2011}, young red giants
\citep{Deheuvels2012} and evolved red giants \citep{Beck2012,
  Mosser2012}. These splittings have in particular been precisely
determined for the red giant KIC~8366239 by
\citet{Beck2012}. \citet{Eggenberger2012} and \citet{Marques2013} have
shown that models including shellular rotation predict a steep
rotation profile which is incompatible with the measured
splittings. \citet{Eggenberger2012} also demonstrated that the
asteroseismic measurements strongly constrain the efficiency of the
needed additional mechanism for the angular momentum transport.

In this work, we study the case of KIC~7341231, a low-mass and
low-metallicity halo early red giant. The very fine observations of
this star by \emph{Kepler} have allowed the precise measurement of
rotational splittings of mixed modes by \citet{Deheuvels2012} who were
able to constrain radial rotation profile of this star by deriving the
core's rotation rate and an upper limit of its surface's. In the
present paper we present a modeling of KIC~7341231 in order to compare
the modeled radial rotation profile with the observed one. In
Section~\ref{KIC}, we present the known characteristics of
KIC~7341231. We then detail the way we modeled the star and discuss
our first results in Section~\ref{Model}, while the influence of
varying various parameters of the model on the obtained radial
differential profile is studied in
Section~\ref{ModParam}. Section~\ref{SolidRot} then presents the
modifications of the modelling in order to mimic a strong angular
momentum transport on the Main Sequence and their consequences on the
rotation deduced and in Section~\ref{SubEv} we focuss on the subgiant
evolution and the different angular momentum transports during this
phase. Finally, our conclusions are given in Section~\ref{Conclu}.


\section{Modeled star: KIC~7341231}
\label{KIC}

KIC~7341231, also known as HIP~92775 and G205-42, is an early red
giant with a mass of about 0.84~$\text{M}_\odot$. Its \emph{Kepler}
magnitude is 9.910. It has an effective temperature comprised between
5470~K \citep{2010A&A...512A..54C} and 5483~K \citep{Ammons2006} and a
log $g$ between 3.55 \citep{Deheuvels2012} and 4.06
\citep{Molenda-Zakowicz2008}. Its metallicity [Fe/H] is very low and
ranges from $-2.18$ \citep{1988AJ.....95.1843L} to $-0.79$
\citep{Ammons2006}. This low metallicity combined with its high proper
motion (39.18~mas/yr in RA and 255.25 in DEC according to
\citealt{2007A&A...474..653V}) and its high radial velocity ($-269.16
\ \text{km} \cdot \text{s} ^{-1}$, \citealt{2002AJ....124.1144L})
indicate that KIC~7341231 is a halo star. All these caracteristics are
summarised in Table~\ref{table1}.

\begin{table*}
  \centering
  \caption{Observational characteristics of KIC 7341231.}
  \begin{tabular}{c x{2.95cm} @{$\ \pm$\ } p{2.95cm} c}
    \hline\hline
    Observables & \multicolumn{2}{c}{Values} & Sources \\
    \hline 
    \multirow{2}{*}{$\text{T}_\text{eff}$} & 5470 & 30 K & Casagrande et al. 2010 \\
     & 5483 & 60 K & Ammons et al. 2006 \\
     \multirow{2}{*}{log $g$} & 3.55 & 0.03 & Deheuvels et al. 2012 \\
     & 4.06 & 0.29 & Molenda-\.Zakowicz et al. 2008 \\
    \multirow{2}{*}{[Fe/H]} & $-2.18$ & 0.06 dex & Laird et al. 1988 \\
     & $-0.79$ & 0.14 dex & Ammons et al. 2006 \\
    \multirow{2}{*}{Proper motion} & 39.18 & 0.85 $\text{mas} \cdot \text{yr}^{-1}$ (RA) &
    \multirow{2}{*}{van Leeuwen 2007} \\
     & 255.55 & 1.24 $\text{mas} \cdot \text{yr}^{-1}$ (DEC) &
      \\
    Radial velocity & $-269.16$ & 0.14 $\text{km} \cdot \text{s}^{-1}$ &
    Latham et al. 2002 \\
    \hline
  \end{tabular}
  \label{table1}
\end{table*}

The radial rotation profile of KIC~7341231 has been derived by
\citet{Deheuvels2012}. They used a \emph{Kepler} short-cadence
photometric light curve of a year long, corrected following the
procedures described by \citet{Garcia2011}. Their analyse of this
light curve resulted first in obtaining the values of the global
seismic parameters: a large separation of $\Delta \nu = 28.9 \pm 0.2\
\mu$Hz and a period spacing (for dipole g-modes) of $\Delta \Pi _1
=107.1 \pm 2.3$~s. They then measured 40 individual eigenmodes of both
acoustic and mixed natures, making possible to study the individual
rotational splittings of these modes. It allowed them to derive the
core's rotation rate and an upper limit for the surface's rotation
rate and thus to constrain the radial rotation profile of the star.
\citet{Deheuvels2012} have shown that the core of KIC~7341231 is
rotating at a frequency $\Omega_\text{c} = 710 \pm 51$~nHz (averaged
on the innermost 1.4\% of the stellar radius, corresponding to 17\% of
the total mass) while its surface rotates at a much slower rate
$\Omega_\text{s} < 150 \ \pm 19$~nHz. These properties are summarised
in Table~\ref{table2}.

\begin{table}
  \centering
  \caption{Seismically derived properties of KIC 7341231 (Deheuvels et
    al. 2012).}
  \begin{tabular}{c r @{$\ \pm$\ } l}
    \hline\hline
    Quantities & \multicolumn{2}{c}{Values} \\
    \hline 
    $\Delta\nu$ & 28.9  & 0.2 $\mu$Hz \\
    $\Delta\Pi_1$ & 112.8 & 0.3 s \\
    $\Omega_c$ & 710 & 51 nHz \\
    $\Omega_s$ & $< 150$ & 19 nHz \\
    \hline
  \end{tabular}
  \label{table2}
\end{table}

KIC~7341231 has then a much lower mass and a much lower metallicity
than the red giant KIC~8366239 observed by \citet{Beck2012}, which has
a mass of about 1.5~$\text{M}_\odot$ and a solar metallicity. It is
then particularly interesting to compare the rotation profile of
KIC~7341231 inferred by observations and the ones predicted by
theoretical models to investigate if the conclusions of
\citet{Eggenberger2012} on KIC~8366239 about the efficiency of the
internal transport of angular momentum in red giants are still valid
for a low-mass, low-metallicity red giant like KIC~7341231, located
near the base of the red giant branch.




\section{Modeling and first results}
\label{Model}

\begin{table}
  \centering
  \caption{Characteristics of the computed models.}
  \begin{tabular}{c r @{$\ \pm$\ } l}
    \hline\hline
    Quantities & \multicolumn{2}{c}{Values} \\
    \hline 
    $\text{[Fe/H]}$ & \multicolumn{2}{c}{$-1\ \text{dex}$}\\
    $\text{Y}_\text{ini}$ & \multicolumn{2}{c}{0.260}\\
    $\text{v}_\text{ini}$ & \multicolumn{2}{c}{$2\ \text{km}\cdot\text{s}^{-1}$}\\
    \hline
    M & 0.84 & 0.01 $\text{M}_\odot$ \\
    Age & 13.01 & 0.06 Gyr \\
    \hline
    $\Delta\nu$ & 30 & 4 $\mu$Hz \\
    $\Delta\Pi_1$ & 115 & 9 s \\
    $\Omega_c$ & 33 & 6 $\mu$Hz \\
    $\Omega_s$ & 36 & 10 nHz \\
    \hline
  \end{tabular}
  \label{table3}
\end{table}

In order to compute models of KIC~7341231, we use the \emph{Geneva
  stellar evolution code} \citep{2008Ap&SS.316...43E} in which the
effects of shellular rotation and the associated meridional
circulation and turbulence on stellar evolution have been implemented
\citep{Eggenberger2010b}. In this work, we do not consider internal
gravity waves or magnetic fields \citep[e.g.][]{Talon2005,
  2005A&A...440..653M}. All computed models start at the Zero Age Main
Sequence (ZAMS). We assume a metallicity $\text{[Fe/H]}=-1 \
\text{dex}$ and an initial helium abundance
$\text{Y}_\text{ini}=0.260$. The stop point we select for the
computed models corresponds to when the models' asymptotic value of the
large separation 
\begin{IEEEeqnarray}{0l}
  \Delta\nu = \left( 2 \int_0^R \frac{dr}{c}\right)^{-1} ,
\end{IEEEeqnarray}
where $c$ is the sound speed,
is equal to the observed one -- without computing
surface corrections. We then compare the models' asymptotic
value of the period spacing
\begin{IEEEeqnarray}{0l}
  \Delta\Pi_1 = \frac{\pi}{\sqrt{2}}\left(\int_{r_1}^{r_2}
    \frac{N}{r}dr\right)^{-1} ,
\end{IEEEeqnarray}
where $N$ is the Brunt-V\"ais\"al\"a frequency
and $\ N(r)>0\ $ for $\ r_1<r<r_2$ , 
to verify that it has the same value as
the observed one. In order to best reproduce the observed surface
rotation, we focuss on models with a low initial rotational velocity
on the ZAMS of $\text{v}_\text{ini}=2 \ \text{km} \cdot \text{s}
^{-1}$.

\begin{figure}
  \centering
  \includegraphics[width=0.45\textwidth]{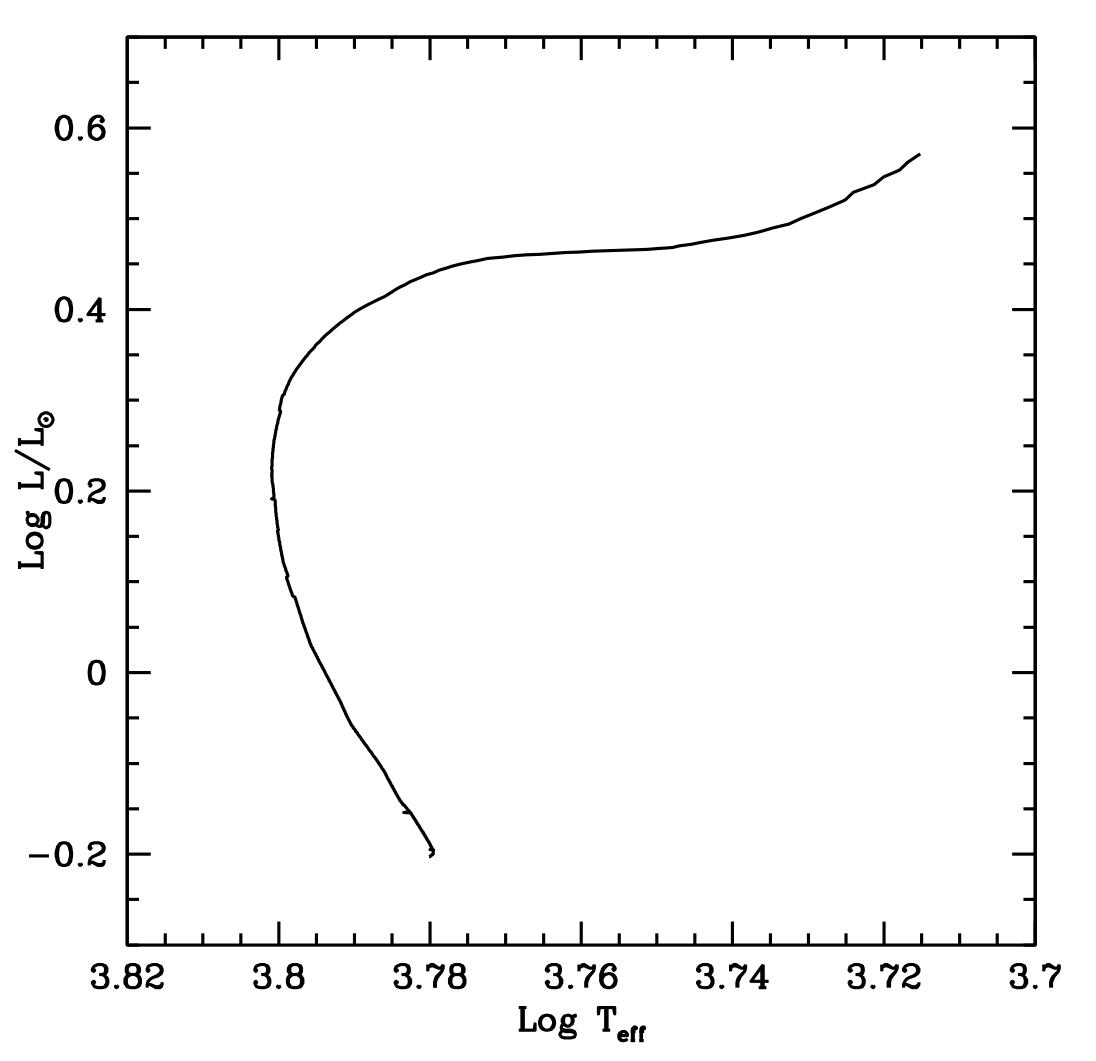}
  \caption{Evolutionary track of the $0.84 \ \text{M}_\odot$ model
    from the ZAMS, including the effects of rotation. The end point
    corresponds to the selected model of the star, matching both the
    observed large separation and period spacing.}
  \label{Evol}
\end{figure}

\begin{figure}
  \centering
  \includegraphics[width=0.45\textwidth]{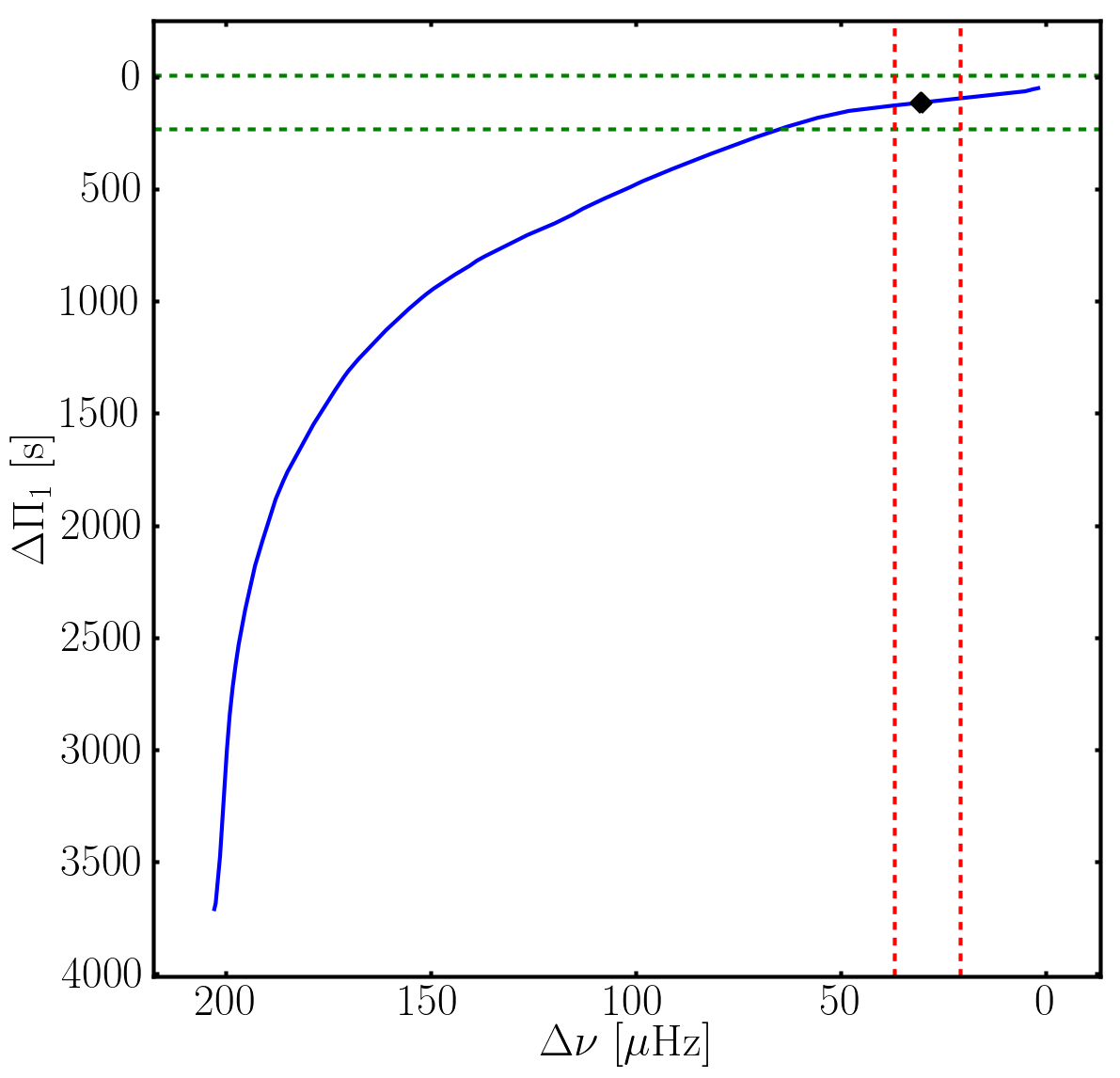}
  \caption{Evolutionary track of the $0.84 \ \text{M}_\odot$ model in
    the asteroseismic H.-R. diagram (blue). The black diamond indicates
    the position of the selected model. The dashed red and green lines
    correspond to the errors bars on the observed large separation and
    period spacing, respectively, from \citet{Deheuvels2012}. These
    error bars have been magnified 
    40 times for the large separation and 400 times for the period
    spacing in order to be visible.}
  \label{dhr_astero}
\end{figure}

In accordance with \citet{Deheuvels2012}, we find that a model with a
mass $\text{M}=0.84 \ \text{M}_\odot$ reproduces both the observed
large separation and period spacing at the same age $\text{T}=13.01 \
\text{Gyr}$. The caracteristics of this model are recalled in
Table~\ref{table3}. Its evolutionary track (Figure~\ref{Evol} and
Figure~\ref{dhr_astero}) shows that KIC~7341231 is an early red giant
that has recently finished the subgiant phase.

\begin{figure}
  \centering 
  \includegraphics[width=0.45\textwidth]{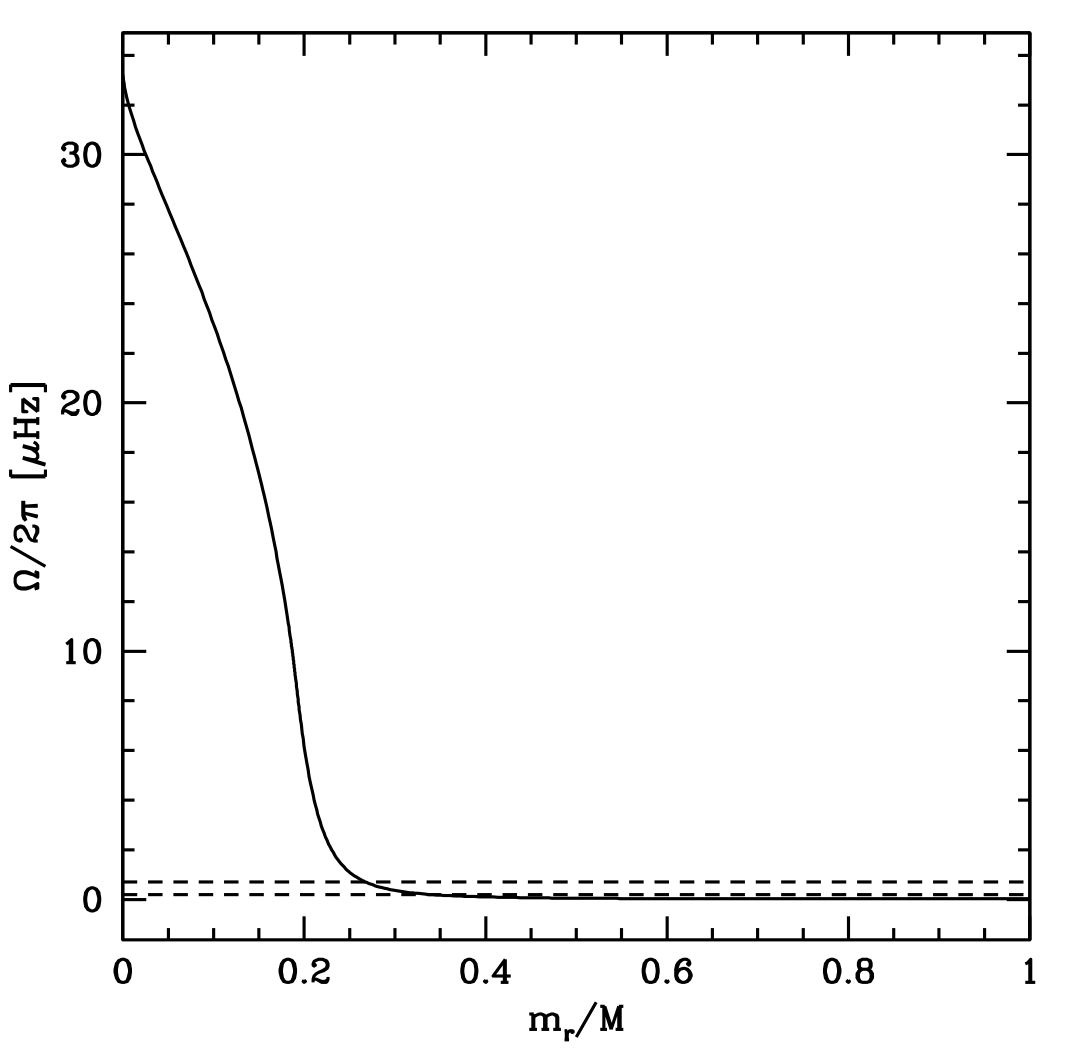}
  \caption{Rotation profile of the selected model at the end of the
    evolutionary track (solid line). The two dashed lines correspond
    to the core and the surface rotation rates derived by \citet{Deheuvels2012}.}
  \label{RotProf1}
\end{figure}

We find that the obtained radial differential rotation profile is
much steeper than the observed one, as can be seen in
Figure~\ref{RotProf1}: the rotation rate of the core is almost two
orders of magnitude higher than the observed value (33~$\mu$Hz for the
model versus 0.71~$\mu$Hz for the observations).This is in complete
agreement with the results obtained by \citet{Eggenberger2012} for the
more massive star KIC~8366239, by \citet{1989ApJ...338..424P} and
\citet{2010ApJ...715.1539T} for the Sun and also by
\citet{Marques2013} for a 1.3~$\text{M}_\odot$ star. It shows that,
for a low-mass and low-metallicity red giant like KIC~7341231, the
shellular rotation and the subsequent meridional circulation and shear
mixing alone produce an internal coupling insufficient to explain the
observed rotation profile.


\section{Influence of modeling parameters}
\label{ModParam}

In order to investigate the effects of varying different modeling
parameters on the rotation profile we compute models with the same
global properties (mass, $\text{v}_\text{ini}$, metallicity\dots) but
with modified properties \citep[see also][]{Marques2013}. We study the
influence of the magnetic braking during the main sequence, the atomic
diffusion, the value of the mixing-length parameter, the value of
initial helium abundance and the metallicity. The resulting
characteristics of these modified models are summarised in 
Table~\ref{table_param}. The reasons why these parameters were chosen
are detailed in the following paragraphs.

\begin{table*}
  \centering
  \caption{Characteristics of the modified models.}
  \begin{tabular}{c c c c r @{$\ \pm$\ } l r @{$\ \pm$\ } l r @{$\
        \pm$\ } l}
    \hline\hline
    Model \# & Modified parameter & Reference value &
    Modified value & \multicolumn{2}{c}{Age (Gyr)} &
    \multicolumn{2}{c}{$\Delta\nu$ ($\mu$Hz)} &
    \multicolumn{2}{c}{$\Delta\Pi_1$ (s)} \\ 
    \hline 
    1 & Magnetic braking & Off & On & 12.97&0.09 & 34&5
    & 122&10 \\
    2 & Atomic diffusion & Off & On & 12.65&0.09 & 32&5
    & 119&10 \\
    3 & $\alpha_\text{MLT}$ & 1.6 & 1.7 & 12.97&0.09 & 34&6
    & 121&10 \\
    4 & $\text{Y}_\text{ini}$ & 0.26 & 0.3 & 9.71&0.07 & 33&6
    & 123&12 \\
    5 & [Fe/H] & - 1 & - 0.8 & 14.16&0.08 & 31&5
    & 116&8 \\
    \hline
  \end{tabular}
  \label{table_param}
\end{table*}

We first wanted to study the influence of magnetic braking on the
modeled rotation profile, as it modifies directly the surface rotation
of the star. Hence, switching on the magnetic braking during the main
sequence reduces the surface rotation rate which amplifies the radial
differential rotation and thus modifies the structure of the star and
increases the strength of the meridional circulation.

We then investigated the effect of atomic diffusion -- or microscopic
diffusion -- on the rotation profile. To calculate the reference
model, we had taken into account only transport mecanisms linked to
the rotation, while here we add atomic diffusion to these transport
mechanisms. This would modify the transport of chemicals in the star,
changing its structure and thus, indirectly, the angular momentum transport.

A parameter that is often fixed to a solar value in stellar
models is the mixing lenght parameter $\alpha_\text{MLT}$. It is
however very difficult to be sure that this parameter can be set to
the same value for a whole range of different stars. Indeed, when
modeling red giants, one can wonder if the vast differences between
the Sun's and red giants' structures cannot prevent us from using the
solar value. That is why we wanted to see the effect of a change in
$\alpha_\text{MLT}$. We thus calculated a model similar to the
reference one in every aspects except for the mixing-length parameter
that we set to 1.7 instead of 1.6. The extension of the convective
zone being increased by this change, the distribution of angular
momentum should be altered as well.

As a halo star, KIC~7341231 is assumed to have a low initial helium
abundance. It is however difficult to have a precise estimate of its
value. In order to evaluate the influence of this quantity on the
internal differential rotation profile, we calculated a model with a
slightly higher $\text{Y}_\text{ini}$: 0.3 instead of 0.26. This model
would correspond to a star born later than KIC~7341231 and could show
the differences in angular momentum transport between two stars formed
at two different epochs. This change has also a strong influence on
the age of the model, which is 9.71~Gyr. While this difference with
the reference model is logical, it is resulting in a higher luminosity
and a slightly higher log~$g$.

For the same reasons why KIC~7341231 is assumed to have a low initial
helium abundance, it is also assumed to have a very low
metallicity. Once again, it remains quite difficult to estimate
accurately this quantity. That is why we decided to assess the
influence of the metallicity on the obtained rotation profile. To do
so, we computed a model with a metallicity of $\text{[Fe/H]}=-0.8$ 
(in accordance with \citealt{Ammons2006}) instead of $-1$. Like the
helium-richer model, this metal-richer model would correspond to a
star formed later than KIC~7341231. Here again, the age of the model
is strongly modified: 14.16~Gyr, far from the 13.01~Gyr of the
reference model. This leads to a higher effective temperature, a lower
luminosity and a slightly more elevated log~$g$.

All these modifications are bound to influence the structure
and evolution of the modeled star, thus resulting in different values
of the large separation and period spacing. Nevertheless, these
changes are relatively small and the obtained values are always
compatible with the observed ones, as can be seen in
Table~\ref{table_param}.

\begin{figure}
  \centering
  \includegraphics[width=0.45\textwidth]{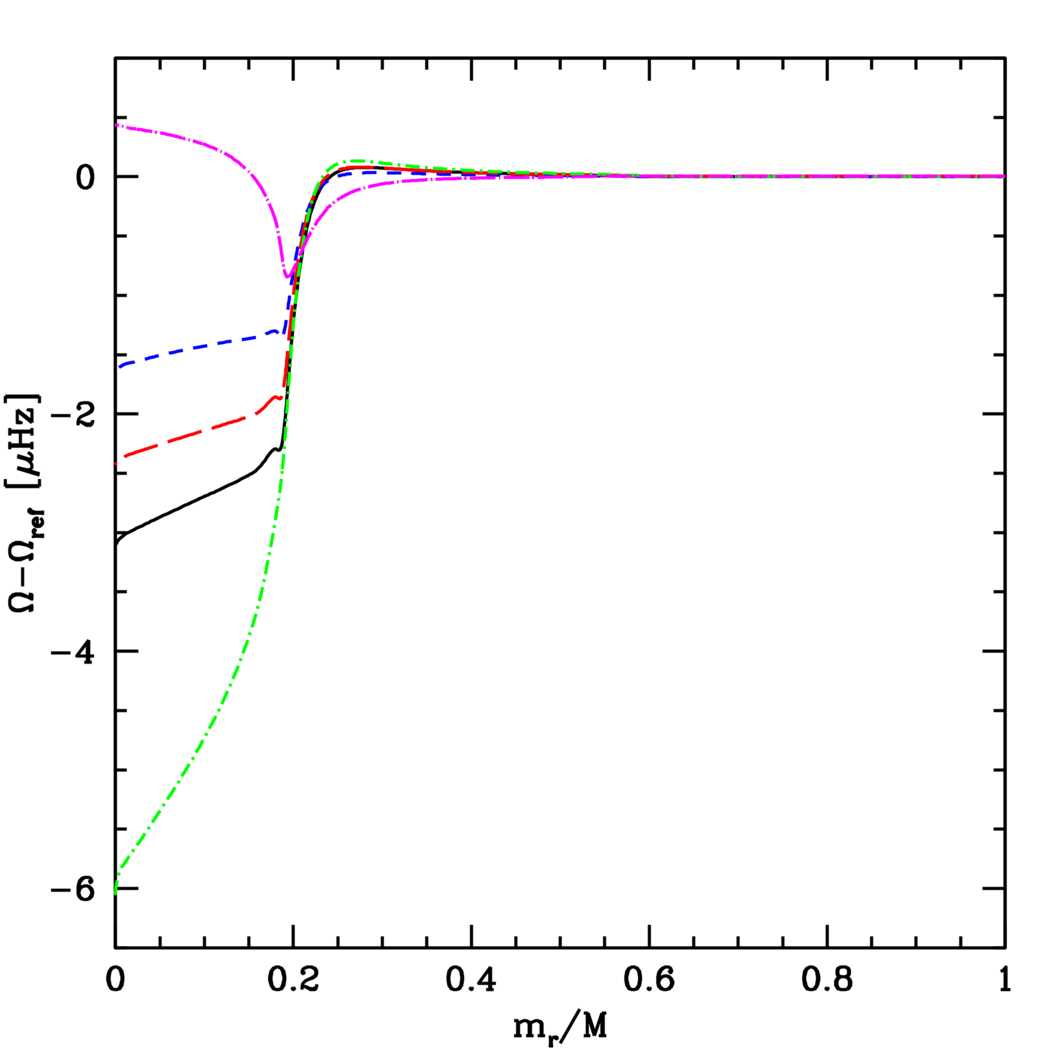}
  \caption{Differences between the rotation profiles of the modified
    models and the rotation profile of the reference model. Black,
    solid: Model 1 (Magnetic braking On). Blue, short-dashed: Model 2
    (Atomic diffusion On). Red, long-dashed: Model 3 (modified
    $\alpha_\text{MLT}$). Green, dot-short-dashed: Model 4 (modified
    $\text{Y}_\text{ini}$). Magenta, dot-long-dashed: Model 5
    (modified [Fe/H]).}
  \label{Comp_rot}
\end{figure}

\begin{table}
  \centering
  \caption{Modifications of the rotation rates of the modified models.}
  \begin{tabular}{c c c}
    \hline\hline
    Model \# & Change for $\Omega_c$ & Change
    for $\Omega_s$ \\ 
    \hline 
    1 & - 9 \% & + 8 \% \\
    2 & - 5 \% & + 8 \% \\
    3 & - 8 \% & + 19 \% \\
    4 & - 18 \% & + 5 \% \\
    5 & + 2 \% & + 5 \% \\
    \hline
  \end{tabular}
  \label{table_change}
\end{table}

The changes implied in the rotation profile are however more
significant. In Figure~\ref{Comp_rot}, we show the difference between
the rotation profiles of the modified models and the reference model's
one. The modifications of the core rotation rate and the surface
rotation rate are also summarised in Table~\ref{table_change}. It is
then clear that modifying the modeling parameters induces small
changes in the rotation profile compared to the gap between the
modeled and observed core rotation rates. We are still far from
reproducing the relatively flat rotation profile deduced from
asteroseimic measurements. Once again, this suggests that another and
yet undetermined physical process is at work in stellar interiors to
increase angular momentum transport, in addition to meridional
circulation and shear mixing.

For the time being, it would seem that the two best candidates for
angular momentum transport from the core to the more external layers
are internal gravity waves excited by the turbulent convective
envelope \citep{ Zahn1997, Talon2005, Talon2008, Mathis2009,
  1993A&A...279..431S} and fossil magnetic field with related MHD
processes \citep{Dough1998, Eggenberger2005, 2005A&A...440..653M,
  2007A&A...474..145Z, 2011A&A...532A..34S, Spada2010}. These two
processes would tend to damp the differential rotation gradient inside
stars and might explain the internal rotation profile of the Sun and
red giants.


\section{Simulating a strong angular momentum transport on the Main Sequence}
\label{SolidRot}

It is then clear that another process is needed to account for the
important angular momentum transport in stellar interiors. While it is still
unclear what this process is precisely but we wanted to study the
possible effects of such a process on the rotation profile of the
early red giant KIC~7341231. What we know is that this
mechanism would flatten the rotation profile of the star. As the
longuest evolutionary phase in stellar life is the main sequence, the
unknown process would probably strongly impact the rotation profile of
the star during this phase. Following this hypothesis, we decided to
consider an extreme case, corresponding to a complete homogenisation
of angular momentum during the main sequence. This homogenisation is
indeed similar to the effects of a very strong and highly efficient
angular momentum transport mechanism.  The efficiency of such an
additional process being unknown on smaller timescales such as the
subgiant phase, we decided to let the star evolve normally as soon as
the end of the main sequence. We are well aware that this approach is
slightly crude and that the real effects of an additional phenomenon
would be more subtle but we chose this methodology as an extreme case,
in order to put constraints on the aformentioned process and have a
first quantitative evaluation of such a process.

\begin{figure}
  \centering
  \includegraphics[width=0.45\textwidth]{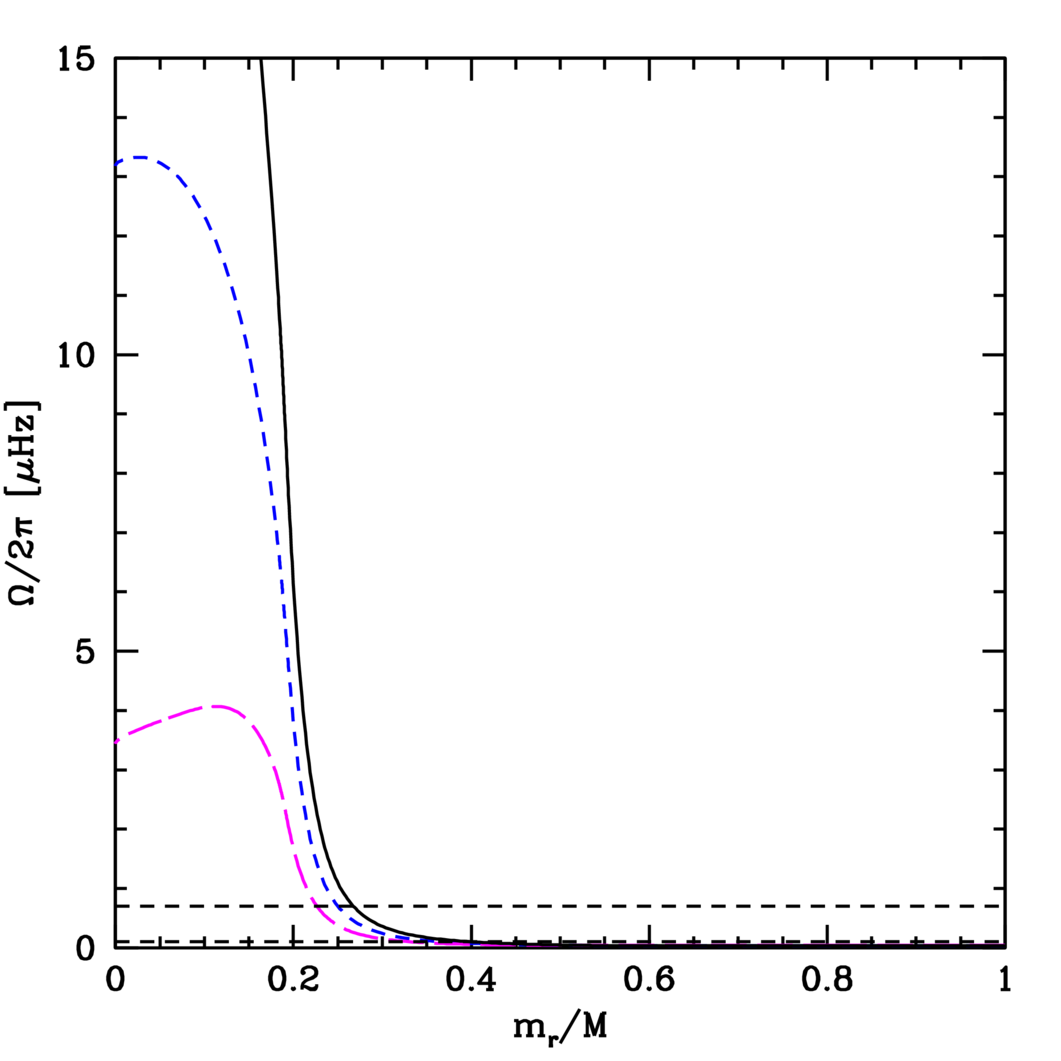}
  \caption{Rotational profiles of the models with and without
    solid-body rotation during the main sequence. Black: same as 
    Fig.\ref{RotProf1}. Blue, short-dashed: model with solid-body rotation until
    $\text{X}_\text{c}=0.1$. Magenta, long-dashed: model with solid-body rotation
    until $\text{X}_\text{c}=0$.} 
  \label{RotProf2}
\end{figure}

In order to simulate a strong homogenizing mechanism, we therefore
forced in the evolution code a solid-body rotation of the star during
the main sequence, allowing the transport of angular momentum through
meridional circulation and shear mixing only during the post-main
sequence evolution. The initial rotational velocity on the ZAMS was
still fixed at $\text{v}_\text{ini}=2 \ \text{km} \cdot \text{s}
^{-1}$. We have computed two such models: for the first one the
solid-body rotation is maintained until the hydrogen abundance in the
core reaches $\text{X}_\text{c}=0.1$, while for the second it stops when
there is no more hydrogen in the core. The first case does not
correspond to the exact end of the main sequence but allowed us to
observe the influence of the rotation of the star during the main
sequence on the red giant's rotation profile. The resulting rotation
profiles (Figure~\ref{RotProf2}) were much less steep than previously,
as could be expected. Without surprise, the later we switch off the
solid-body rotation, the flatter the rotation profile. This shows the
sensitivity of the red giant rotation profile to the efficiency of the
angular momentum transport during the main sequence. Nevertheless, the
gap is still huge between what we obtain and what is observed : the
models' core rotation rates are at least one order of magnitude higher
than the observed one ($\Omega_c=$ 3.5 and 13.2 $\mu$Hz for models
versus 0.71 $\mu$Hz for observations).  This suggests that the
required process, whatever what it might be, has to be efficient also
on small time scales such as the subgiant evolution.

\begin{figure}
  \centering
  \includegraphics[width=0.45\textwidth]{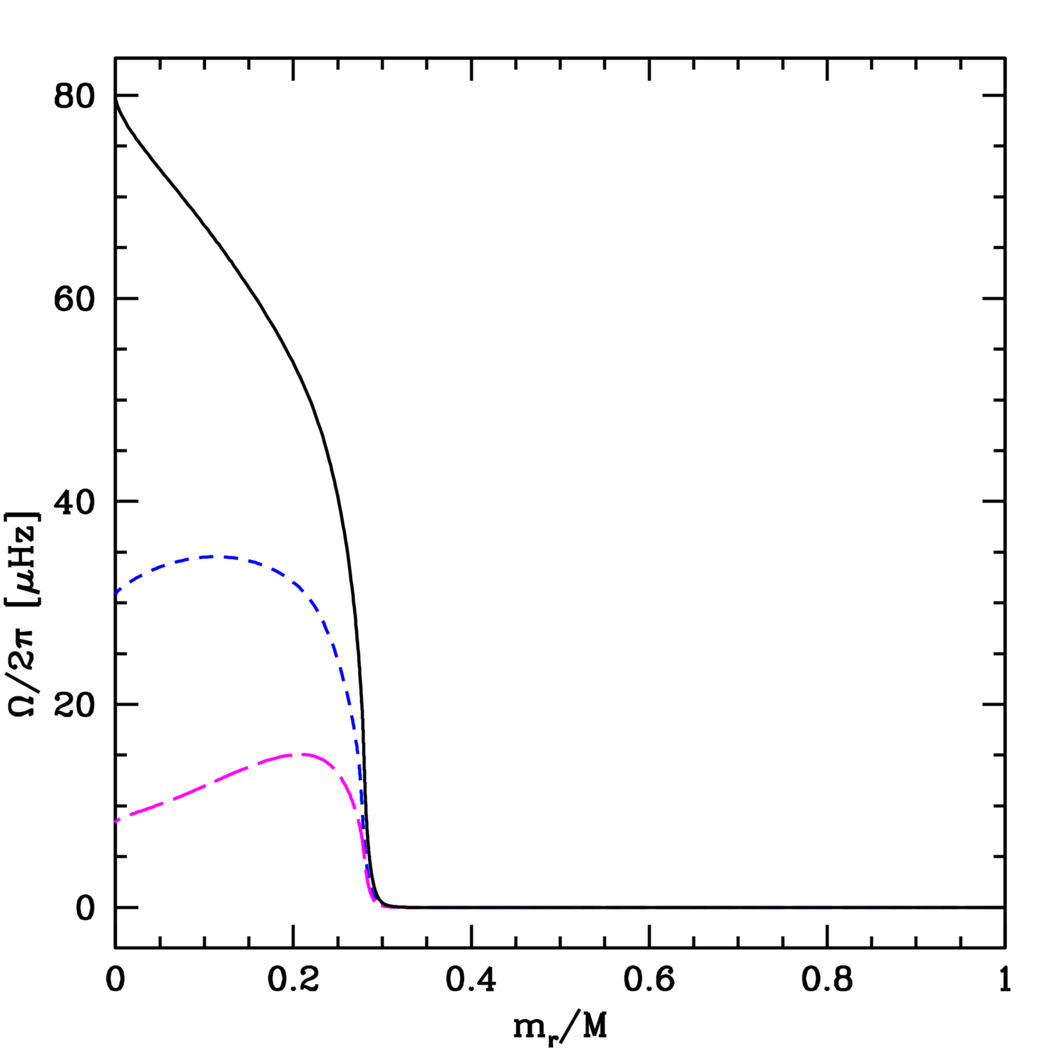}
  \caption{Rotational profiles of the models with and without
    solid-body rotation during the main sequence, but after the
    subgiant phase (at an age $\text{T}=13.40\ \text{Gyr}$). Black:
    normal case (no solid-body rotation). Blue, short-dashed: model
    with solid-body rotation until $\text{X}_\text{c}=0.1$. Magenta,
    long-dashed: model with solid-body rotation until
    $\text{X}_\text{c}=0$.} 
  \label{RotProf3}
\end{figure}

Another remarkable consequence of the solid-body rotation during the
main sequence is that even in the later evolution of the star
(i.e. the red giant phase) the rotation profiles obtained strongly
differ from the one corresponding to the normal case (i.e without
solid-body rotation). This discrepancy can be seen in
Figure~\ref{RotProf3}. In particular, we can observe a decrease of the
rotational rate in the core of the star, which was already visible in
Figure~\ref{RotProf2} but less prominent. This shows that, in these
models, the rotational history of the star is not completely erased by
its evolution during the subgiant and red giant phases. In order to
fully understand red giants' rotation profiles it is thus adamant to
study the evolution of the rotation inside the star during its whole
evolution.


\section{Subgiant evolution and angular momentum transport}
\label{SubEv}

\begin{figure}
  \centering
  \includegraphics[width=0.45\textwidth]{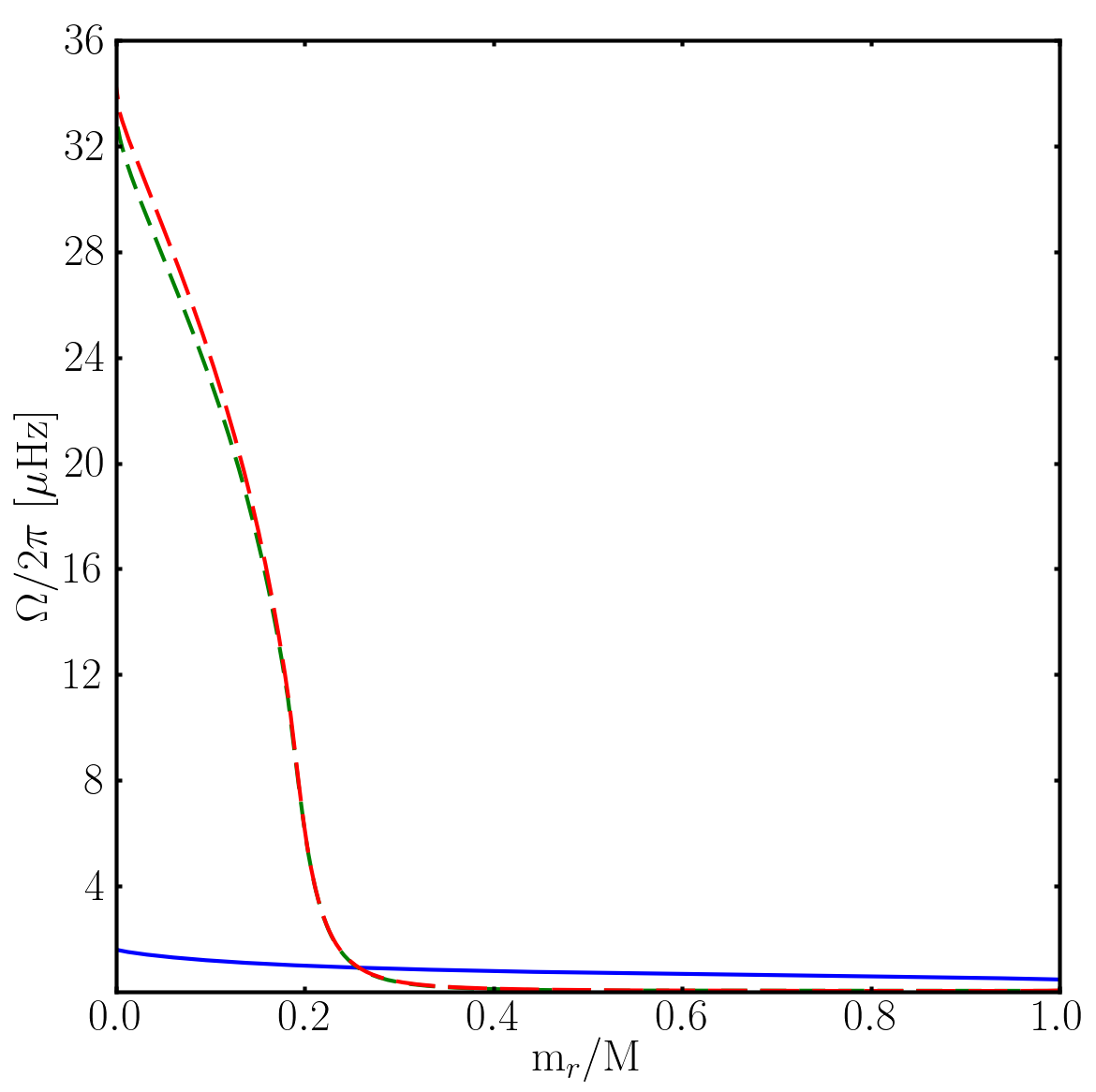}
  \caption{Rotational profiles of the reference model at different
    evolution times. Blue, solid: end of Main Sequence. Green,
    short-dashed: end of Subgiant evolution considering all angular
    momentum transport phenomenons. Red, long-dashed: end of subgiant
    evolution considering only conservation of angular momentum during
    the subgiant phase.} 
  \label{ConsMom1}
\end{figure}

\begin{figure}
  \centering
  \includegraphics[width=0.45\textwidth]{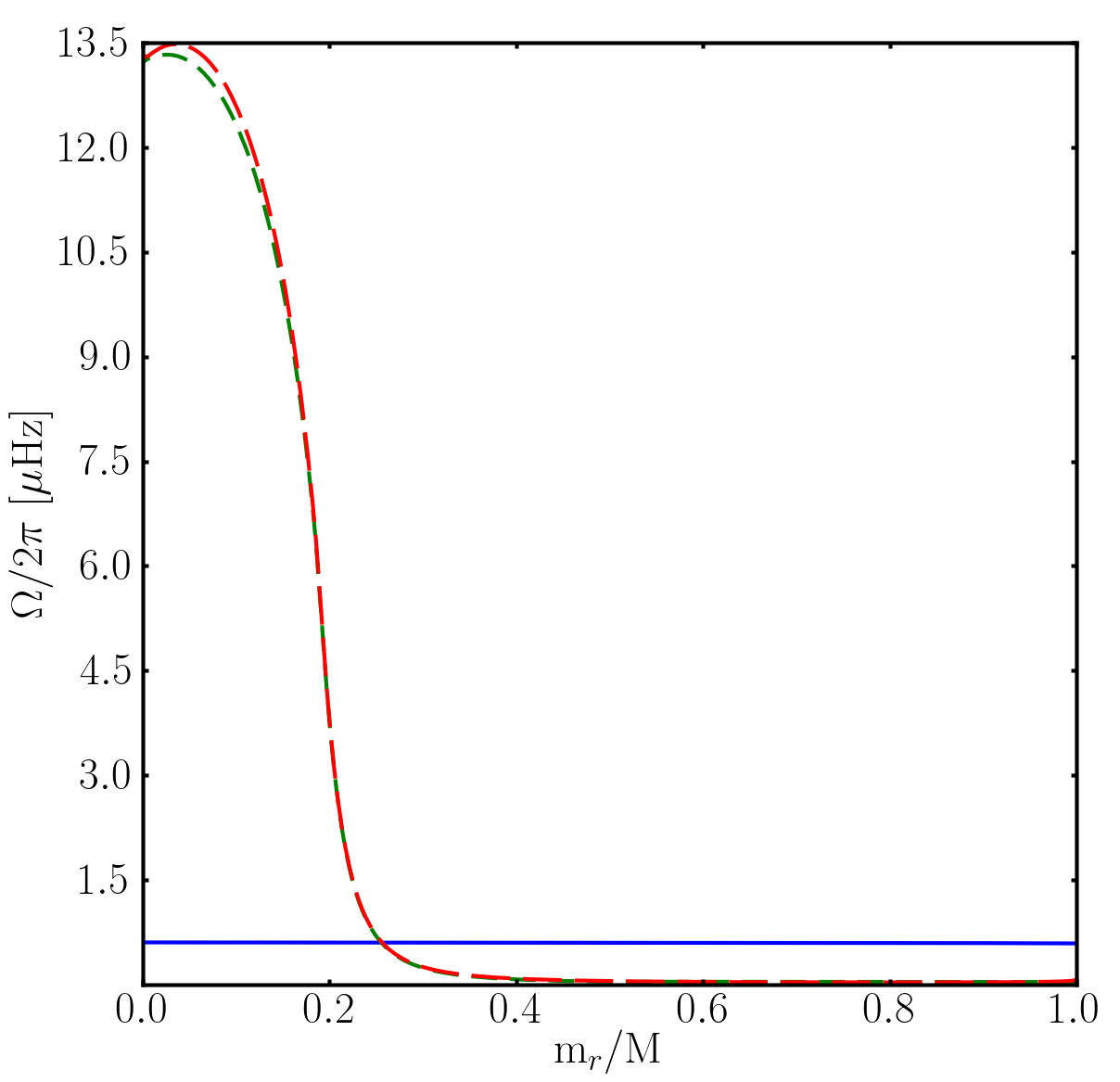}
  \caption{Rotational profiles of the model with solid-body rotation
    during Main Sequence (until $\text{X}_\text{c}=0.1$) at different
    evolution times. Colors: same as Figure~\ref{ConsMom1}.}
  \label{ConsMom2}
\end{figure}

We have seen that the missing angular momentum transport mechanism has to be
efficient during the short-lived subgiant phase. But what of the transport
processes already considered? In order to quantify their effect
during this phase, we compared the evolution of the rotational profile
during the subgiant phase in two different cases. For the first one,
we took into account rotationaly induced processes such as meridional
circulation and shear mixing. For the second one, we considered only
angular momentum conservation. Thus the difference between the
obtained rotational profiles at the end of the subgiant phase is a
good proxy of the efficiency of the transport processes.

This has been done for the reference model and for the model with
solid-body rotation during the main sequence until
$\text{X}_\text{c}=0.1$. The resulting profiles are shown in
Figures~\ref{ConsMom1} and \ref{ConsMom2}. While considering
rotationaly induced transport reduces indeed the core's rotation rate,
the effect is of very small magnitude (1.5~$\mu$Hz maximum compared to
$\Omega_c=33\mu$Hz, a 4.5\% change). The evolution of the star's
rotation profile during subgiant evolution is therefore completely
dominated by the sheer conservation of angular momentum due to the
profound modifications of the star's structure.

This comparison emphasizes that any angular momentum transport
phenomenon considered must be considerably more efficient than
meridional circulation and shear mixing on small timescales. More
precisely, during the subgiant phase, this phenomenon has to be strong
enough to counterbalance angular momentum conservation. Once again,
this is a rather important constraint for the studied additional
processes such as internal gravity waves and magnetic fields.


\section{Conclusion}
\label{Conclu}

The comparison between the radial rotation profile of the red giant
KIC~7341231 inferred from asteroseismic measurements with profiles
predicted by stellar models -- including shellular rotation,
meridional circulation and shear mixing -- have shown that such models
are unable to reproduce the observed internal rotational rates. The
theoretical rotation profiles are indeed too steep compared to the one
deduced from observations: in the case of KIC~7341231 the model's core
is rotating at least 30 times faster than what has been deduced
through asteroseismic means. This is in perfect agreement with the
conclusions of \citet{Eggenberger2012}, \citet{Marques2013} and
\citet{Ceillier2012} and shows that this discrepancy is also found for
a low-mass, low-metallicity red giant like KIC~7341231 \citep[see
also][]{Palacios2006}.

These results illustrate the need to progress in our understanding and
modeling of (magneto-)hydrodynamical processes at work in stellar
interiors and to implement new physical processes into stellar
evolution codes, such as internal gravity waves or magnetic fields.
We have demonstrated here that these processes, however efficient
during the long-lasting main sequence, have to be very effective
during the rapid evolution of the subgiant phase. Furthermore, we have
also shown that their effect on internal rotation must be considered
over the whole evolution of the star and that the rapid evolution
during the subgiant phase does not completely erase all the previous
rotational history of the star.

It would be interesting to determine the efficiency of the unknown
physical process needed for the internal transport of angular momentum
in order to correctly reproduce the observed rotation profile of
KIC~7341231. This would allow a comparison with the effective viscosity
of $3\cdot 10^4\ \text{cm}^2\text{s}^{-1}$ found for the more massive
red giant KIC~8366239 \citep{Eggenberger2012}, although we would like
to remind the reader that neither internal gravity waves nor fossil
magnetic fields have an effect comparable with an effective viscosity
\citep[see][]{2013LNP...865...23M}.

We have no doubt that codes in which the effects on angular momentum
transport of processes like internal gravity waves or magnetic fields
have been implemented will allow us to understand more accurately the
various physical mechanisms at work in stars. In such improvement of
our vision of the dynamical evolution of stars, asteroseismology is
then one of the most promising ways to get strong constraints as
demonstrated here with the case of KIC~7341231.

\begin{acknowledgements}
  TC, RAG and SM acknoledge the CNES support of CoRoT and of
  asteroseismic activities at the SAp -- CEA/Saclay and the CNRS/INSU
  PNPS support. TC and SM thank the Geneva Observatory for its
  hospitality. PE was partly supported by the Swiss National Science
  Foundation. The authors would like to thank the referee for her/his
  suggestions which helped improving the paper.
\end{acknowledgements}

\bibliography{lib_article.bib}

\end{document}